\documentstyle[aps]{revtex}
\input epsf
\title{\bf Quantum group invariant, nonextensive quantum
statistical mechanics}
\author {Marcelo R. Ubriaco\thanks{Electronic address:ubriaco@ltp.upr.clu.edu}\\
Laboratory of Theoretical Physics\\
 Department of Physics\\
 University of Puerto Rico\\
 P. O. Box 23343, R\'{\i}o Piedras\\
 PR 00931-3343, USA}
\begin{document} 
\vspace{0.3in}
\maketitle
\vspace{0.15in}
\begin{abstract}
We  study the consequences of introducing quantum group invariance
in the formalism of nonextensive quantum statistical mechanics. 
 We find that the corresponding thermodynamical system
is equivalent to a Bose-Einstein gas in the Boltzmann-Gibbs formalism
with a higher critical temperature than the standard Bose-Einstein case.
\end{abstract}
\vspace{0.3in}

\section{Introduction}
Since the formulation of nonextensive statistical mechanics \cite{T1}
 an impressive amount of related works has been published in several fields of physics,
mathematics and biology \cite{T2}.  
This formalism is based on the generalized entropy
\begin{equation}
S_q=\frac{k}{q-1}\left(1-\sum_Rp^q_R\right),\label{Sq}
\end{equation}
where $p_R$ is the probability for the ensemble to
be in the state $R$ and $q$ is  parameter which in
principle can be any real number.
The Tsallis entropy $S_q$ possesses the properties of
the Shanon's entropy except that of additivity. Given two
independent systems
$\Sigma$ and $\Sigma'$, the entropy $S_q$ satisfies
the pseudoadditivity property
\begin{equation}
\frac{S_q^{\Sigma\cup\Sigma'}}{k}=\frac{S_q^{\Sigma}}{k}+
\frac{S_q^{\Sigma'}}{k}+(1-q) \frac{S_q^{\Sigma}}{k}\frac{S_q^{\Sigma'}}{k} .
\end{equation}
In addition, in several articles different
authors have addressed the possibility that quantum groups could play a role
in nonextensive physics \cite{T3}. Although nonextensive statistical mechanics
does not embody any quantum group structure,  it shares \cite{A,J}
 with quantum groups the mathematical formalism of $q$-analysis.
As shown in Ref.\cite{A}, the Tsallis entropy can be defined in terms of
a $q$ derivative acting on the probability distribution. This observation
opens the possibility that nonextensive statistics could be also studied
with the use of  mathematical subjects as basic hypergeometric functions
and series \cite{E}, and the theory of partitions \cite{An}.
However, in order to have a quantum group structure we need to include
in a formalism or model a set of either operators or coordinates
satisfying an algebra covariant under some kind of quantum group transformations \cite {WZ,Wo}.
This new structure introduces additional degrees of freedom, 
such that defining a quantum group invariant hamiltonian 
we can  study the consequences that result from this new symmetry \cite{U2}.
A recent study \cite{U4} showed that the thermodynamical behavior
of a gas in the factorization approximation
of nonextensive statistics differs considerably 
with the properties of a quantum group gas in the Boltzmann-Gibbs formulation.
Therefore, it is our interest to study the simplest quantum group invariant nonextensive system.
The thermodynamical system that we will study depends, in principle, on two parameters.
One parameter is related to the generalization of Boltzmann-Gibbs thermodynamics
to the formalism proposed by Tsallis.  The quantum group parameter, relates
the classical Lie group to the quantum group symmetry. Thus, loosely speaking, 
our density matrix is the result of a double deformation of the standard
density matrix.
This paper is organized as follows. In Sec. \ref{2} we  introduce
a quantum group invariant density matrix and calculate the particle
distribution for this model.  With use of consistency arguments we
show how the two parameters are related and their values restricted 
to a certain interval.  In particular, we find that this
system has a Bose-Einstein particle distribution with
a higher critical temperature than  a standard Bose-Einstein gas.
In Sec. \ref{3} we summarize our results.
\section{Nonextensive statistical mechanics and quantum group invariance}\label{2}
In this section we study the consequences of introducing
quantum group invariance in nonextensive statistical mechanics.
The density matrix $\hat{\rho}$ for nonextensive thermodynamics
was reformulated in \cite{TMP,LMR} in terms of normalized $q$-expectation values.
The recent formalism defines the expectation values of an operator as
\begin{equation}
A_q=\frac{Tr\hat{A}\hat{\rho^q}}{Tr\hat{\rho}^q},
\end{equation}
where the density matrix in the grand canonical ensemble is given by the
expression
\begin{equation}
\hat{\rho}=\frac{1}{Z}\left[1-(1-q)\beta^*\left((\hat{H}-U_q)-\mu(\hat{N}-N_q)\right)\right]^{1/(1-q)},
\end{equation}
with the inverse temperature $\beta^*=\beta/c$ and the normalization $c=Tr\hat{\rho}^q$.
The temperature $T_*=c/k\beta$ has been recently shown \cite{AMPP} to be the correct, physical, temperature 
in nonextensive thermodynamics.
We wish to write the operator $\hat{H}-\mu\hat{N}$ in terms of
 quantum operators such that it is invariant under unitary quantum group transformations.
The form of this operator can be inferred from the observation \cite{U2} that
 operators $\Phi_{\alpha}$ and their adjoints $\overline{\Phi}_{\alpha}$
satisfying  the quantum group  algebra $SU_{q'^{1/2}}(M)$ are defined by the equations
\begin{eqnarray}
\Phi_{\alpha}&=&\phi_{\alpha}^{\dagger -1}q'^{\sum_{\lambda=\alpha+1}^M N_{\lambda}/2}\{N_{\alpha}\},\\
\overline{\Phi}_{\alpha}&=&\phi_{\alpha}^{\dagger}q'^{\sum_{\lambda=\alpha+1}^M N_{\lambda}/2},
\end{eqnarray}
where $\phi_{\alpha}$ and $\phi_{\alpha}^{\dagger}$ are standard boson operators, $N_{\alpha}$
is the number operator and the bracket
denote
the $q'$-number $\{x\}=(1-q'^x)/(1-q')$. A simple algebraic manipulation
 shows that the simplest quantum group invariant operator
 $h=\sum_{\alpha}\overline{\Phi}_{\alpha}\Phi_{\alpha}$,
can be written in terms of the $q'$-number operator as follows
\begin{equation}
h=\{\sum_{\alpha=1}^M N_{\alpha}\}.\label{h}
\end{equation}
In our case, in addition to the sum over the internal (quantum group) degrees of freedom
we have also a summation over the momentum.  Therefore, based on the 
form of the operator $h$ in Eq. ({\ref{h})  we propose the following operator
\begin{equation}
\hat{H}-\mu\hat{N}=\Lambda\left\{\sum_{\kappa}\sum_{\alpha=1}^{M}\beta^*\epsilon'_{\kappa}
\hat{N}_{\kappa,\alpha}\right\},
\end{equation}
where $M$ is the number of internal degrees of freedom, $\epsilon'_{\kappa}=\epsilon_{\kappa}-\mu$,
$q'$ is related to the quantum group parameter and $\Lambda$ is a constant to be determined.
The number operator is written more explicitly as $\hat{N}_{\kappa,\alpha}=\phi^{\dagger}_{\kappa,\alpha}
\phi_{\kappa,\alpha}$ where the operators $\phi_{\kappa,\alpha}$ and its adjoint are
standard boson operators. A simple inspection shows that the operator
$\hat{H}-\mu\hat{N}$ can be written as a summation over the internal degrees of freedom as follows
\begin{equation}
\hat{H}-\mu\hat{N}=\Lambda\sum_{\alpha=1}^M\left\{\sum_{\kappa}\beta^*\epsilon'_{\kappa}\hat{N}_{\kappa,\alpha}\right\}
q'^{\sum_{\lambda=\alpha+1}^M\sum_{\kappa'}\beta^*\epsilon'_{\kappa'}\hat{N}_{\kappa',\lambda}}.\label{H}
\end{equation}
Since in the $q'\rightarrow 1$ limit the Eq. (\ref{H}) should become the standard expression 
 $\hat{H}-\mu\hat{N}=\sum_{\alpha}\sum_{\kappa}\epsilon'_{\kappa}\hat{N}_{\kappa,\alpha}$,
we expect that the factor $\Lambda$ to be a linear function of the   temperature $1/\beta^*$.
A representation of the quantum group operators can be obtained from the observation that
  Eq. (\ref{H})  rewritten
as a product $\sum_{\alpha}\overline{\Phi}\Phi$ define the operators $\Phi$ and the adjoints as 
\begin{eqnarray}
\overline{\Phi}_{\kappa,\alpha}&=&\phi^{\dagger}_{\kappa,\alpha}
q'^{\sum_{\lambda=\alpha+1}^M\sum_{\kappa'}(\beta^*/2)\epsilon'_{\kappa'}\hat{N}_{\kappa',\lambda}} \nonumber\\
\Phi_{\kappa,\alpha}&=&\phi^{\dagger-1}_{\kappa,\alpha} \label{Phi}
q'^{\sum_{\lambda=\alpha+1}^M\sum_{\kappa'}(\beta^*/2)\epsilon'_{\kappa'}\hat{N}_{\kappa',\lambda}} 
\left\{\sum_{\kappa''}\beta^*\epsilon'_{\kappa''}\hat{N}_{\kappa'',\alpha}\right\}.
\end{eqnarray}
In particular, for $M=2$ and after redefinition of the operators $\Phi$ and $\overline{\Phi}$ as
\begin{eqnarray}
\Omega_{\kappa,\alpha}&=&\frac{1}{\{\beta^*\epsilon'_{\kappa}\}^{1/2}}\Phi_{\kappa,\alpha}\nonumber ,\\
\overline{\Omega}_{\kappa,\alpha}&=&\frac{1}{\{\beta^*\epsilon'_{\kappa}\}^{1/2}}\overline{\Phi}_{\kappa,\alpha},
\end{eqnarray}
it is simple to show that the operators $\Omega$ satisfy the 
$SU_{Q_{\kappa}}(2)$ covariant algebra
\begin{eqnarray}
\Omega_{\kappa,2}\overline{\Omega}_{\kappa,2}&=& 1+Q_{\kappa}^2
\overline{\Omega}_{\kappa,2}\Omega_{\kappa,2}\nonumber\\
\overline{\Omega}_{\kappa,1}\Omega_{\kappa,2}&=&Q_{\kappa}^{-1}\Omega_{\kappa,2}
\overline{\Omega}_{\kappa,1}\nonumber\\ \label{SU}
\Omega_{\kappa,1}\Omega_{\kappa,2}&=&Q_{\kappa}^{-1} \Omega_{\kappa,2}\Omega_{\kappa,1}\nonumber\\
\Omega_{\kappa,1}\overline{\Omega}_{\kappa,1}&=&1+Q_{\kappa}^2
\overline{\Omega}_{\kappa,1}\Omega_{\kappa,1}-(1-Q_{\kappa}^2)\overline{\Omega}_{\kappa,2}
\Omega_{\kappa,2},
\end{eqnarray}
where $Q_{\kappa}=q'^{\beta^*\epsilon'_{\kappa}/2}$.
The equations (\ref{SU})  are covariant under the field redefinitions $\Omega'=T\Omega$ and 
$\overline{\Omega'}=\overline{\Omega}\;\overline{T}$ where the quantum matrix $T$ is a $2\times 2$ matrix
given by
\begin{equation}
T=\left(\begin{array}{cc} a & b  \\
 c & d \end{array}\right),
\end{equation}
with the elements $\{a,b,c,d\}$ generating the algebra
\begin{eqnarray}
ab=Q_{\kappa}^{-1}ba   & , &\;\;\;\; ac=Q_{\kappa}^{-1}ca \nonumber \\
bc=cb & , & \;\;\;\;dc=Q_{\kappa}cd  \nonumber \\
db=Q_{\kappa}bd & , & \;\;\;\;\; da-ad=(Q-Q_{\kappa}^{-1})bc  \nonumber \\
& & det_{Q_{\kappa}}T\equiv ad-Q_{\kappa}^{-1}bc=1 .
\end{eqnarray} 
The matrix $\overline{T}$ is obtained from the unitary conditions \cite{VWZ} $\overline{a}=d, \overline{b}=q^{-1}c$
and $q\in {\bf R}$. As is well known, the $SU_{Q_{\kappa}}(2)$ transformation $T$ matrix and the corresponding $R$ matrix 

$$R=\left(\begin{array}{cccc} Q_{\kappa} & 0 & 0 & 0 \\ 0 & 1 & 0 & 0 \\
0 & Q_{\kappa}-Q_{\kappa}^{-1} & 1 & 0 \\ 0 & 0 & 0 & Q_{\kappa}\end{array}\right)$$

satisfy the algebraic relations \cite{Ta}
\begin{equation}
RT_1T_2=T_2T_1R,\label{T},
\end{equation}
and
\begin{equation}
R_{12}R_{13}R_{23}=R_{23}R_{13}R_{12},
\end{equation}
where $T_1=T\otimes 1$ and $T_2=1\otimes T$
$\in V\otimes V$, $(R_{23})_{ijk,i'j'k'}=
\delta_{ii'} R_{jk,j'k'} \in V\otimes V\otimes V$.
 We should remark that quantum group covariance of
the operator algebra is satisfied for those operators $\Omega$ with the same 
momentum $\kappa$. For example, for $\kappa\neq \kappa'$ we have the relation
\begin{equation}
\overline{\Omega}_{\kappa',1}\overline{\Omega}_{\kappa,2}=Q_{\kappa}\overline{\Omega}_{\kappa,2}
\overline{\Omega}_{\kappa',1},
\end{equation}
which is not quantum group covariant \cite{U1}. For $M>2$, and hereafter omitting
the $\kappa$ index, the Eq. (\ref{SU}) 
are generalized to the set of equations
\begin{equation}
\Omega_{\alpha}\overline{\Omega}_{\lambda}=\delta_{\alpha\lambda}+ Q R_{\eta\lambda\alpha\gamma}
\overline{\Omega}_{\gamma}\Omega_{\eta}\label{c1}
\end{equation}
\begin{equation}
\Omega_{\alpha}\Omega_{\lambda}=Q^{-1}R_{\gamma\eta\lambda\alpha}\Omega_{\gamma}\Omega_{\eta},\label{c2}
\end{equation}
The $M^2\times M^2$ matrix 
 $R_{\alpha\lambda\gamma\eta}$ is well known in the quantum group literature \cite{WZ}, and in terms of
our parameter $Q_{\kappa}$ is explicitly written 
\begin{equation}
R_{\alpha\lambda\gamma\eta}=\delta_{\alpha\gamma}\delta_{\lambda\eta}(1+(Q-1)\delta_{\alpha\lambda})
+(Q-Q^{-1})\delta_{\lambda\gamma}\delta_{\alpha\eta}\theta(\alpha-\lambda),
\end{equation}
where $\theta(\alpha -\lambda)=1$ for $\alpha >\lambda$ and zero otherwise.
From Eqs. (\ref{H}) and (\ref{Phi}) we see that the quantum group invariant density matrix can be simply written as
\begin{equation}
\hat{\rho}=\frac{1}{Z}\left[1-(1-q)\beta^*\left(\sum_{\alpha=1}^M \Lambda \overline{\Phi}_{\alpha}\Phi_{\alpha}+\mu N_q-U_q)\right)\right]^{1/(1-q)}.
\label{rho1}
\end{equation}
 Eq.  (\ref{rho1}) is the simplest quantum group invariant density matrix
one can write. It represents, after all, the density matrix  for a set of free quantum group bosons $\Phi$.
By choosing the constant $\Lambda$ to have the value
\begin{equation}
\Lambda=\frac{1-q'}{(1-q)\beta^*}-(1-q')(\mu N_q-U_q),
\end{equation}
we find that the density matrix satisfies a simple relation with boson operators
\begin{equation}
\hat{\rho}^q \phi^{\dagger}_{\kappa,\alpha}=\phi^{\dagger}_{\kappa,\alpha} q'^{q\beta^*\epsilon'_{\kappa}/(1-q)}
\hat{\rho}^q. \label{rho}
\end{equation}
Since other than integer powers of the operators $\phi$ and $\phi^{\dagger}$ are meaningless,
the operator $\hat{\rho}^q$  only makes sense for those values of  the nonextensive parameter $q$  such that
$q/(1-q)$ is a positive integer.  Therefore, the parameter is restricted to the values $q=n/n+1$,for $n=1,2,3...$.
The density matrix in Eq. (\ref{rho}) depends on two parameters.  The parameter $Q_{\kappa}$ labels, for a given value
of $\kappa$,  the deformation from a classical $SU(M)$  to a quantum  $SU_{Q_{\kappa}}(M)$ symmetry.
The parameter $q$ indicates the departure from the Boltzmann-Gibbs formalism to nonextensive 
thermodynamics. A simple inspection of the operator $\hat{H}-\mu\hat{N}$ for $q'\rightarrow 1$
shows that it has the correct limit $\sum_{\alpha}\sum_{\kappa}\epsilon'_{\kappa} N_{\kappa,\alpha}$
 if the two parameters are related according to
\begin{equation}
1-q'=(1-q)f(q),\label{f}
\end{equation}
where $f(q)\rightarrow 1$ as $q\rightarrow 1$. This relation implies that we cannot
take the two limits $q'\rightarrow 1$ and $q\rightarrow 1$ independently without
avoiding the trivial solution $f(q)=0$. In addition, according to Eq. (\ref{f}) the quantum group parameter
$q'$ is also restricted to the interval $0<q'\leq 1$.
 We consider for simplicity and without loss of generality the case $q=q'$. The
average number of particles with energy  $\epsilon_{\kappa}$ is given by
\begin{eqnarray}
n_{\kappa}&=&Tr\hat{\rho}^q\phi^{\dagger}_{\kappa}\phi_{\kappa}/Tr\hat{\rho}^q\nonumber\\
&=&\frac{M}{q^{-q\beta^*\epsilon'_{\kappa}/(1-q)}-1},\label{n}
\end{eqnarray}
which is functionally equivalent to the Bose-Einstein distribution in Boltzmann-Gibbs
thermodynamics. It is simple to check that the correct limit of the
Bose-Einstein distribution $n_{\kappa}=1/(exp(\beta\epsilon'_{\kappa})-1)$ is obtained for $q=1$.
Therefore, the particle distribution is equivalent to the Bose-Einstein
textbook case 
\begin{equation}
n_{\kappa}=\frac{M}{e^{\beta'^*(\epsilon_{\kappa}-\mu)}-1},\label{n'}
\end{equation}
where  $\beta'^*=-\beta^*q\ln q/(1-q)$ and
the fugacity $z=exp(\beta'^*\mu)$.  With use of some algebra we find
that the normalization constant $c=Tr \hat{\rho}^q$ is given by the equation
\begin{equation}
c=\frac{1}{Z^q}Tr\left[q^{\sum_{\kappa}\sum_{\alpha=1}^{M}\beta^*\epsilon'_{\kappa}\hat{N}_{\kappa,\alpha}}
\left(1-(1-q)\beta^*(\mu N_q-U_q)\right)\right]^{q/(1-q)},
\end{equation}
leading to 
\begin{equation}
c=\frac{1}{Z^q}\left[1-(1-q)\beta^*(\mu N_q-U_q)\right]^{q/(1-q)}\prod_{\kappa}\frac{1}
{1-q^{q\beta^*M\epsilon'_{\kappa}/(1-q)}}.
\end{equation}
Taking into account that the constant $c=Z^{1-q}$ \cite{LMR} we obtain that
\begin{equation}
\ln c=q \ln\left[1-(1-q)\beta^*(\mu N_q-U_q)\right]-
(1-q)\sum_{\kappa}\ln\left(1-q^{q\beta^*M\epsilon'_{\kappa}/(1-q)}\right).
\end{equation}
The summation over $\kappa$ can be approximated, as usual, by an integral with the result
\begin{equation}
c=\left[1-(1-q)\beta^*(\mu N_q-U_q)\right]^q 
exp\left[\frac{-4 V(1-q)^{5/2}}{(qM|\ln q|)^{3/2}\sqrt{\pi}\lambda^{*3}}g_{5/2}(z^M)\right],\label{c}
\end{equation}
where the Bose-Einstein function $g_{5/2}(x)=\sum_{n=1}x^n/n^{5/2}$ and the thermal wavelength
$\lambda^*=h\sqrt{\beta^*/2\pi m}$.
Since the particle distribution  in Eq. (\ref{n'}) is of the Bose-Einstein type this
system will exhibit Bose-Einstein condensation but with a different critical temperature
than the textbook case. 

The results for the total  average number of particles $N_q$
and the total energy $U_q=Tr\sum_{\kappa}\sum_{\alpha}\epsilon_{\kappa}\phi^{\dagger}_{\kappa}
\phi_{\kappa}\hat{\rho}^q/c$, can be readily copied 
from the standard results with the replacement of $\beta$  by $\beta'^*$
\begin{eqnarray}
N_q&=&N_0+\frac{MV}{\lambda'^{*3}}g_{3/2}(z),\\
U_q&=&\frac{3}{2}\frac{MV}{\beta'^*\lambda'^{*3}}g_{5/2}(z),
\end{eqnarray}
with $\lambda'^*=h\sqrt{\beta'^*/2\pi m}$.  The critical temperature
is therefore given by the expression
\begin{equation}
T_c^*=\frac{-(1-q)}{q\ln q}T_c^{BE},\label{t}
\end{equation}
where $T_c^{BE}$ is the critical temperature for a Bose-Einstein
gas.
Since the factor in Eq. (\ref{t}) is $\frac{-(1-q)}{q\ln q}>1$    we find that for the 
interval $0<q<1$
the critical temperature  satisfies that $T_c^*>T_c^{BE}$.
Figure 1 is a graph that shows the dependence of the critical temperature $T^*_c$ on the
parameter $q$.\\
\epsfxsize=400pt \epsfbox{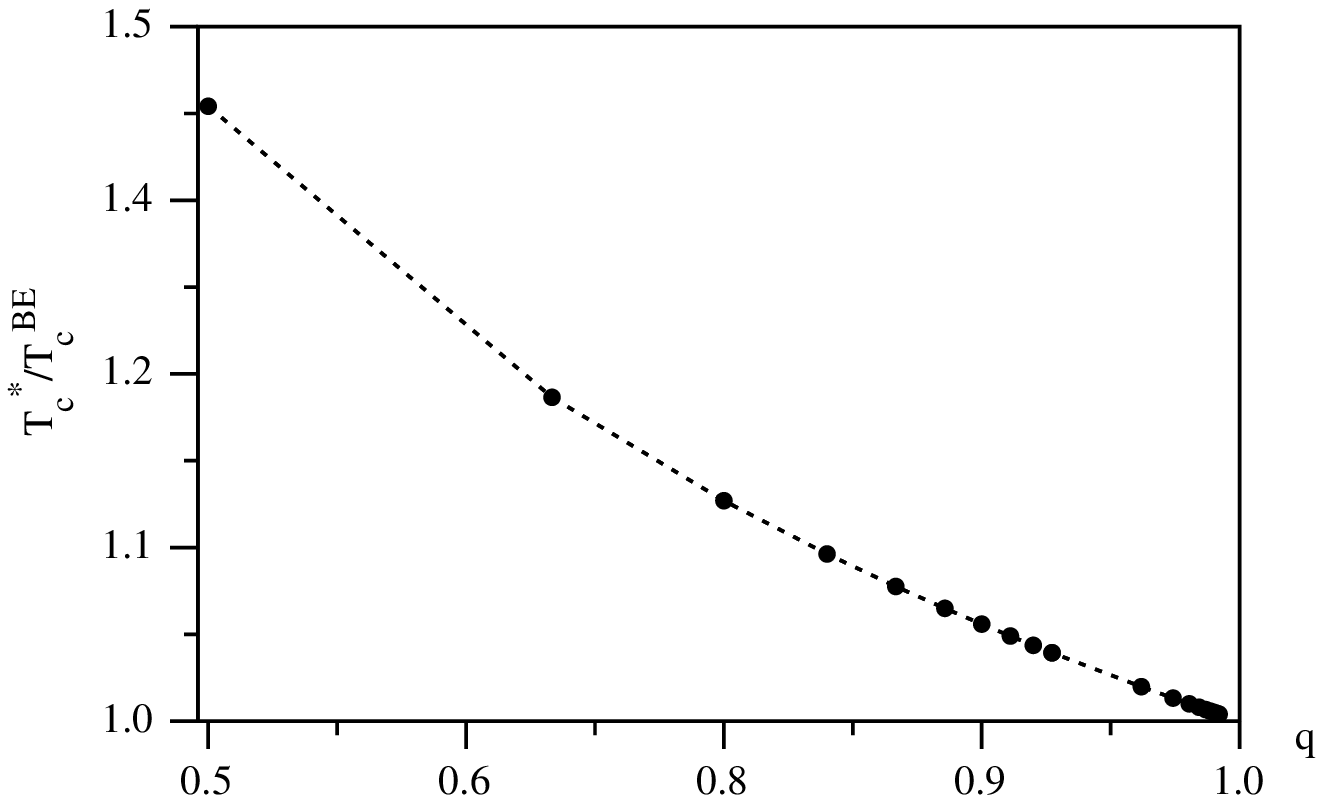}\\
{\footnotesize FIG. 1. The dependence of $T_c^*/T_c^{BE}$ on the parameter $q$ where $T_c^*$ is
the critical temperature of our thermodynamic system and  $T_c^{BE}$
is the critical temperature for $q=1$. The dots indicate the allowed values of $q$} .

\section{Conclusions}\label{3}

In this paper we have been able to define a quantum group invariant density matrix
in nonextensive quantum statistical mechanics.  For a given momentum $\kappa$
 the quantum group operators  satisfy algebraic relations covariant 
under $SU_{q'^{\beta^*\epsilon'_{\kappa}/2}}(M)$ transformations.  
Although the
two deformations in the density matrix are originally unrelated, we found that the
the parameters $q'$ and $q$ are related through the equation $q'=1-(1-q)f(q)$, where
$f(q)$ is an arbitrary function satisfying $f(q=1)=1$ and the nonextensive parameter $q$
takes values $q=n/n+1, n=1,2,3,...$  A simple calculation showed
that the particle distribution function 
 is equivalent to the standard Bose-Einstein distribution but the critical temperature
is higher than the critical temperature of the standard case.  Thus, according to
the results reported in this paper, the thermodynamics of a quantum group
invariant system in nonextensive thermodynamics is equivalent to the thermodynamical
behavior of a system of free bosons in the extensive, Boltzmann-Gibbs, formulation.


\begin{thebibliography}{99}
\bibitem{T2} A list of works can be found in tsallis.cat.cbpf.br/biblio.htm.
\end{thebibliography}
\end{document}